\documentclass[aps,prc,amsfonts,amssymb,twocolumn,floatfix]{revtex4}
\usepackage{graphicx}
\usepackage{bm}

\begin{document}

\title{Extrapolation Method for the No-Core Shell Model}

\date{\today}

\author{H. Zhan}
\author{A. Nogga}
\author{B. R. Barrett}
\affiliation{Department of Physics, University of Arizona, Tucson, 
Arizona 85721}

\author{J. P. Vary}
\affiliation{Department of Physics and Astronomy,
Iowa State University, Ames, Iowa 50011}

\author{P. Navr\'{a}til}
\affiliation{Lawrence Livermore National Laboratory, L-414, P.O. Box
  808, Livermore, California 94551}

\begin{abstract}
Nuclear many-body calculations are computationally demanding. 
An estimate of their accuracy is often hampered by the limited 
amount of computational resources even on present-day supercomputers. 
We provide an extrapolation method based on perturbation theory, 
so that the binding energy
of a large basis-space calculation can be estimated without diagonalizing
the Hamiltonian in this space. The extrapolation method is tested 
for $^3$H  and $^6$Li nuclei. It will extend our computational 
abilities significantly and allow for reliable error estimates.
\end{abstract}


\maketitle

\section{Introduction}

In recent years a great deal of progress has been made in solving the
nuclear many-body problem based on microscopic nuclear
interactions. Using a stochastic approach 
\cite{pieper01b,suzukilec,koonin97} 
or directly diagonalizing the Hamiltonian \cite{navratil02},
previous investigations have been able to improve 
the accuracy of the calculations to a level, so that
rather small parts of the Hamiltonian, like the 
three-nucleon (3N) forces, can be probed by a comparison of the 
predicted spectra to the experimental values. 
For unambiguous conclusions, a reliable error bound 
for the calculations is of highest importance. 
Currently, error bounds have often been established
using benchmark calculations \cite{kamada01c,barrett03,navratil02,
navratil00b}.
In this paper, we propose a new scheme to extend no-core shell model (NCSM)
calculations beyond their current limits, and to estimate error bounds for
existing calculations. 
We establish a correlation between the expectation value 
of the Hamiltonian $\langle H \rangle$ with respect to the approximate 
ground state and that of the square of 
the Hamiltonian 
$\langle H^2\rangle$, following the ideas of Mizusaki and Imada 
\cite{mizusaki02,mizusaki03}. 
This will improve the estimate of binding energies and provide more 
reliable error estimates. 

To this aim, we proceed with a brief introduction to the main concepts
of NCSM calculations in Section~\ref{sec:ncsm}. Then we motivate 
the correlation by the properties of NCSM effective interactions and 
perturbation theory in Section~\ref{sec:scaling}.
A comparison between the extrapolated results and the known
binding energies for $^3$H and $^6$Li 
is given in Section~\ref{sec:extrap}. 
Conclusions and outlooks to the applications of this method 
and a discussion of the 
relation of this work to previous ones in 
Refs.~\cite{mizusaki02,mizusaki03} are presented 
in Section~\ref{sec:concl}.

\section{No-core shell model}
\label{sec:ncsm}
Usually shell-model calculations assume an
inert nuclear core. Taking only the valence nucleons as active
particles clearly has the advantage of reducing  the number of many-nucleon
states. But, so far, this also means that effective
shell model interactions have to be used, which cannot simply be related 
 to nucleon-nucleon (NN) interactions, as they have been developed for
few-nucleon systems. 

The NCSM approach is different. All
nucleons are taken to be active, and the same interactions
are used as they are in traditional few-body methods. The calculations 
are performed in a finite antisymmetrized harmonic oscillator (HO) basis,
using either Jacobi \cite{navratil00a} or Cartesian coordinates like those 
in usual shell model investigations. In general, the short range repulsion
of nuclear forces cannot be easily described in a finite basis. In
particular, the HO basis is not well suited to describe the short range 
correlations and, on the other hand, the exponential tail of bound-state 
wave functions. Consequently, effective interactions appropriate to the 
basis-size truncation must be derived from the underlying nuclear forces
in order to achieve convergence with a manageable number of basis 
states. These effective interactions can be systematically related to the 
``bare'' NN interactions \cite{suzuki80,suzuki82,navratil00b,navratil00c}.
The scheme described in Refs.~\cite{suzuki82,navratil00b,navratil00c} 
is based on unitary 
transformations of the Hamiltonian \cite{okubo54}, 
which decouple the model space from the complete Hilbert space 
describing the quantum mechanical system. 
  
The starting
point of all NCSM calculations is a non-relativistic  $A$-body Hamiltonian,
which includes two-body interactions. The extension to 3N forces
has been introduced in \cite{navratil03,mncb02}. We do not take them 
into account in this study. 
Adding the center-of-mass (CM) HO potential, which is subtracted at a 
later stage in the calculation, one can cast 
the Hamiltonian \cite{lipkin58}
into  the form
\begin{eqnarray}
\label{eq:omham}
H_A^\Omega & = & \sum_{i=1}^A \left ({ {\vec p_i}^2 \over 2 m_i } + { 
 \ m_i \ \Omega^2 \over 2 } \ {\vec {r_i}}^2  \right)  \cr
& & +  \sum_{i<j=1}^A \left( V_{ij} - { m_i \ m_j \over 2 M_A} \ \Omega^2
 (\vec r_i - \vec r_j) ^2 \right) ,
\end{eqnarray}
where $\vec p_i$ ($\vec r_i$) is the momentum (position) of the 
particle $i$ with mass $m_i$, and  $M_A=\sum_i m_i$ is the total mass 
of the $A$ particles.

It is possible to establish a unitary transformation
of the Hamiltonian, which decouples two parts of the Hilbert space -- a
rather small finite model space $P$ and the rest of the Hilbert space
$Q$. The projection operators on the two spaces are also called $P$ and
$Q$. They fulfill the relations $Q=1-P$, and 
$Q e^{-S} H_A^\Omega e^{S} P = 0$, where $e^{S}$ is the unitary 
transformation \cite{navratil00b}.

The solution of the $A$-body problem is made possible by the observation that,
for nuclear problems, a cluster approximation to the full unitary
transformation sufficiently speeds up the convergence compared to the bare 
interactions, so that practical calculations are possible. 
Instead of solving the
full $A$-body problem, the procedure is carried out for a much smaller
$a$-body problem, with $a=2$ or $a=3$ in practice. For such cases, one 
starts with the Hamiltonian 
\begin{eqnarray}
\label{eq:clusterham}
H_a^\Omega & = & \sum_{i=1}^a \left ({ {\vec p_i}^2 \over 2 m_i } + { 
 \ m_i \ \Omega^2 \over 2 } \ {\vec {r_i}}^2  \right)  \cr
& & +  \sum_{i<j=1}^a \left( V_{ij} - { m_i \ m_j \over 2 M_A} \ \Omega^2
 (\vec r_i - \vec r_j) ^2 \right) .
\end{eqnarray}
Note that the mass of the full $A$-body system enters into the strength
constant of the relative HO potential. Therefore, for this $a$-body 
Hamiltonian, the HO
interaction does not cancel in the relative motion, and provides a confining
mean field interaction. However, the HO interaction is cancelled in the 
$A$-body calculation. This procedure improves the convergence of our 
$A$-body results with increasing model space size.

Using the unitary transformation satisfying $Q_a e^{-S_a} H_a^\Omega
e^{S_a} P_a = 0$, one determines an effective Hamiltonian 
$H^\Omega_{eff,a}=P_a e^{-S_a} H_a^\Omega e^{S_a} P_a$,
which exactly describes the $a$-body cluster in the model space. 
The truncation $P_a$ from the full $a$-body Hilbert space is related to $P$
by the requirement that the $a$-body states included in $P$ are also 
included in $P_a$. One then
defines the effective interaction to be used for the $A$-body calculation as 
\begin{equation}
V_{eff,a}^{\Omega} = H_{eff,a}^\Omega - \sum_{i=1}^a  \left ({ {\vec p_i}^2 \over 2 m_i } + { 
 \ m_i \ \Omega^2 \over 2 } \ {\vec {r_i}}^2  \right)  .
\end{equation}
With increasing size of the model space,
the effective interaction converges to the
``bare'' interaction, so that the ``bare'' problem is recovered, meaning 
that the approximation is controllable. 

Due to the cluster approximation, for $a < A$ we no longer have an exact
effective interaction. This shows up, for example, in an $\Omega$
dependence of the binding energy. 
However, experience shows that shell-model
calculations converge much faster, if performed with effective forces, as
defined above \cite{navratil02}. 
All calculations in this paper are based on effective interactions obtained 
from two-body cluster solutions.

\section{Correlation between $E_0$ and $\Delta E$} 
\label{sec:scaling}

The binding energy $E_0=\langle H \rangle$ evaluated in an approximate 
ground state must approach the exact binding energy $\mathcal{E}_0$ as
the energy variance 
$\Delta E^2 = \langle H^2 \rangle - \langle H \rangle^2$ vanishes. 
One can easily extrapolate the exact binding energy with a series of 
approximate calculations, once the behavior of $E_0$ as a function of 
$\Delta E$ is determined. It is suggested by Mizusaki and Imada that 
$E_0$ can be expanded in terms of $\Delta E^2 E_0^{-2}$ for small values of 
$\Delta E^2 E_0^{-2}$ \cite{mizusaki02, mizusaki03}. They propose two 
extrapolation formulae for traditional shell model calculations:
$E_0 \simeq a\,\Delta E^2 E_0^{-2} + \mathcal{E}_0$ and
$E_0 \simeq a_0\,\Delta E^2 + a_1\Delta E^4 + \mathcal{E}_0$,
where $a$, $a_0$, and $a_1$ are fitting parameters. This has motivated 
us to search for the correlation between  $E_0$ and $\Delta E$ in context 
of the NCSM, and establish an extrapolation method for the NCSM.

\subsection{Notation}
\label{sec:notation}

For the following investigation we need to explicitly define the model spaces. 
To this aim,
we truncate the full Hilbert space spanned by the antisymmetrized HO basis
at a maximum total HO quantum number $N_m$. This $N_m$ counts the number 
of oscillator quanta including the lowest oscillator configuration for the 
nucleus of interest\footnote{The quantum number, $N_m$, differs from 
``$N_{max}$'' of 
Refs.~\cite{navratil00b,navratil00c} by the oscillator quanta of the 
lowest unperturbed oscillator configuration. Thus $N_m=N_{max}$ for $^3$H,
and $N_m = N_{max} + 2$ for $^6$Li.}.
This truncation ensures that all states are included up to 
a given energy, so that spurious CM motion can be projected out.
The goal of the
following study is to acquire results for a fixed $N_m$ from
calculations for even smaller subspaces $\widetilde P$ of $P$ truncated by 
$\widetilde N_m \le  N_m$. The effective interactions will be
obtained for the larger $N_m$ in all cases. 

We now switch to matrix notation, because it helps to 
demonstrate the structure of the effective Hamiltonian, 
$\mathbf{H}$, which can be decomposed into
\begin{equation} \label{eq:bb}
\mathbf{H} = \left[ \begin{array}{ll}
\widetilde{\mathbf{H}} & \mathbf{B}\\ 
\mathbf{B}^\mathrm{T} & \widehat{\mathbf{H}}
\end{array} \right] = \mathbf{H}_0 + \mathbf{H}_1,
\end{equation}
where 
\begin{equation}
\mathbf{H}_0 = \left[ \begin{array}{ll}
\widetilde{\mathbf{H}} & \mathbf{0}\\ 
\mathbf{0} & \widehat{\mathbf{H}}
\end{array} \right], \mbox{ and } 
\mathbf{H}_1 = \left[ \begin{array}{ll}
\mathbf{0} & \mathbf{B}\\ 
\mathbf{B}^\mathrm{T} & \mathbf{0}
\end{array} \right] . 
\end{equation}

The boundaries of the blocks in $\mathbf{H}$
are given by the truncation $\widetilde N_m$ of the subspaces $\widetilde P$. 
$\widetilde{\mathbf{H}}$ acts only in $\widetilde P$ and 
$\widehat{\mathbf{H}}$ in the remainder space $P- \widetilde P$. 
Because the effective interactions and $\mathbf{H}$ 
are defined for the complete $P$ space,
$\widetilde{N}_m$ completely determines $\widetilde{\mathbf{H}}$, 
so one may write explicitly 
$\widetilde{\mathbf{H}}(\widetilde{N}_m)$. We assume that $\bm{\psi}$
and $\bm{\phi}$ are the eigenvectors of $\widetilde{\mathbf{H}}$ and 
$\widehat{\mathbf{H}}$, respectively, i.e., 
$\widetilde{\mathbf{H}}\psi_i = E_i\psi_i, \ i = 0\ldots \tilde{n}-1$,
and $\widehat{\mathbf{H}}\phi_j = U_j\phi_j, \ j = 0\ldots \hat{n}-1$, 
where 
$\tilde n$ and $\hat n$ are the dimensions of $\widetilde P$ and $P-\widetilde P$, 
respectively. 
Then the eigenvectors of $\mathbf{H}_0$ are 
\begin{equation}
\mathbf{A} = \left[ \begin{array}{ll} 
\bm{\psi} & \mathbf{0} \\ \mathbf{0} & \bm{\phi} 
\end{array} \right]
\end{equation}
because of the form of $\mathbf{H}_0$. 
The ground state of $\mathbf{H}_0$, i.e., $A_0$, is of the from 
$[\psi_0^\mathrm{T},\mathbf{0}]^\mathrm{T}$, and we denote the ground-state 
energy by $E_0$.

The energy dispersion $\Delta E$ is defined as
\begin{equation} \label{eq:dES}
\Delta E^2 = A_0^{\rm{T}}\mathbf{H}^2 A_0 - 
             (A_0^\mathrm{T} \mathbf{H} A_0)^2
           = \psi_0^{\rm{T}}\mathbf{BB}^{\rm{T}}\psi_0,
\end{equation}
where we have made use of the specific forms of $\mathbf{H}_0$
and $\mathbf{H}_1$. The quantity $\Delta E^2$ 
measures how well $A_0$ approximates the ground state of $\mathbf{H}$.
As $\widetilde{N}_m$ approaches $N_m$, $\Delta E$ will vanish, and 
$E_0$ will become the true binding energy $\mathcal{E}_0$.

To establish a scaling between $E_0$ and  $\Delta E^2$, we try to improve 
the results of $\widetilde{\mathbf{H}}$ using perturbation theory. 
Several more approximations will be necessary to motivate the scaling. 
These approximations will  
be tested numerically in model calculations for the $^3$H binding
energy. We do not intend to improve on the currently available  $^3$H binding
energy results \cite{marsden02,nogga02b,nogga03a,pieper01b}, 
but restrict ourselves to the $P$ space truncated at $N_m=20$, 
which yields realistic, but not fully converged results.
Here we will only aim to recover $P$-space results with
$\widetilde{N}_m<20$ calculations. Note that this is in contrast to the
usual NCSM calculations where one investigates the dependence on $N_m$ 
itself, and the effective interactions are renormalized for each $N_m$.
This also means that our results will explicitly depend on $\Omega$.
But the small model space sizes used will allow us to extract 
intermediate results, which
will demonstrate the origin of the scaling behavior.  
We will also show results for $^6$Li to demonstrate 
the applicability to more complex systems.

\subsection{Application of Perturbation Theory}

\begin{figure}
\centering
\includegraphics[width = 3.2in]{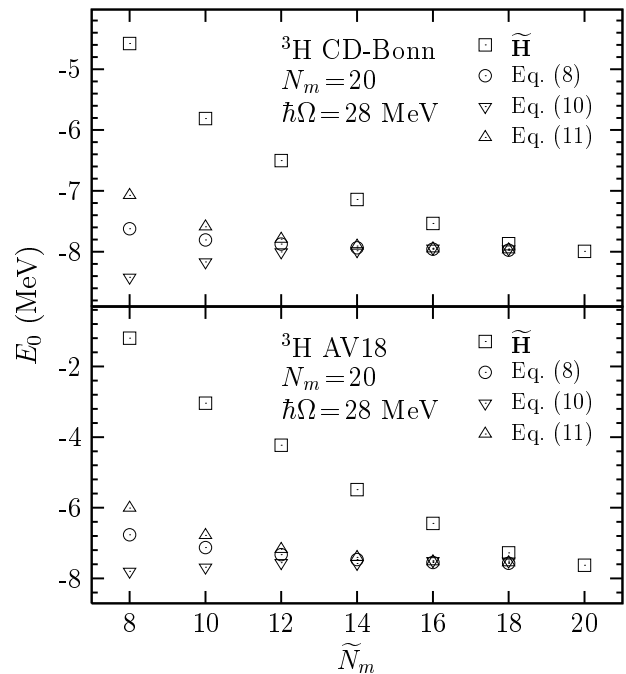}
\caption{Perturbative calculations of the binding energy $E_0$ of
the triton with $N_m=20$ and $\hbar\Omega=28$ MeV. The values of 
$E_0$ for $\widetilde{\mathbf{H}}$ are denoted by open squares.
The circles are calculated using Eq.~(\ref{eq:EiPT}), 
the downward triangles Eq.~(\ref{eq:Ei2PT}), and the upward triangles
Eq.~(\ref{eq:Ei3PT}).
\label{fig:aspt}}
\end{figure}

With the block-block decomposition Eq.~(\ref{eq:bb}), second-order 
perturbation theory gives the eigenenergy of $\mathbf{H}$, i.e.,
$\mathcal{E}_i$, corresponding to $E_i$, as
\begin{equation} \label{eq:EiPT}
\mathcal{E}_i \simeq  E_i + \sum_{j = 0}^{\hat{n}-1}
\frac{\left( \psi_i^\mathrm{T} \mathbf{B} \phi_j \right)^2}{E_i - U_j},
\quad i = 0 \ldots \tilde{n}-1.
\end{equation}
The next non-vanishing perturbative term is of the 4th order. Despite
its simplicity and accuracy (see Fig.~\ref{fig:aspt}), 
Eq.~(\ref{eq:EiPT}) is not useful in establishing a correlation between 
$E_0$ and $\Delta E$, because it requires the knowledge of all the 
eigenvalues and eigenvectors of $\widehat{\mathbf{H}}$.

One may avoid diagonalizing 
$\widehat{\mathbf{H}}$ by a redefinition of the  
decomposition:
\begin{equation}
\mathbf{H}_0 = \left[ \begin{array}{ll}
\widetilde{\mathbf{H}} & \mathbf{0}\\ 
\mathbf{0} & \mathbf{D}
\end{array} \right], \mbox{ and } 
\mathbf{H}_1 = \left[ \begin{array}{ll}
\mathbf{0} & \mathbf{B}\\ 
\mathbf{B}^\mathrm{T} & \mathbf{C}
\end{array} \right],
\end{equation}
where $\mathbf{D}$ is a diagonal matrix containing the diagonal elements of 
$\widehat{\mathbf{H}}$, and
$\mathbf{C} = \widehat{\mathbf{H}} - \mathbf{D}$. 
For NCSM Hamiltonians, this decomposition is appropriate (see subsection 
\ref{sec:scalpert}). We note that Eq.~(\ref{eq:dES}) remains unchanged.
Similar to 
Eq.~(\ref{eq:EiPT}), second-order perturbation theory gives
\begin{equation} \label{eq:Ei2PT}
\mathcal{E}^{(2)}_i =  E_i + \sum_{j = 0}^{\hat{n}-1}
\frac{\left( \psi_i^\mathrm{T} B_j \right)^2}{E_i - D_{jj}},
\quad i = 0 \ldots \tilde{n}-1,
\end{equation}
where $B_j$ is the $j$-th column 
of $\mathbf{B}$. Up to the third order, we have
\begin{equation} \label{eq:Ei3PT}
\mathcal{E}_i \simeq \mathcal{E}^{(2)}_i  +
\sum_{j,l=0}^{\hat{n}-1} \frac{\left(\psi_i^\mathrm{T} B_j\right) 
C_{jl} \left(B_l^\mathrm{T} \psi_i\right)}
{(E_i - D_{jj})(E_i - D_{ll})}.
\end{equation} 

Figure \ref{fig:aspt} shows the binding energy of $^3$H
calculated using equations (\ref{eq:EiPT}), (\ref{eq:Ei2PT}), 
and (\ref{eq:Ei3PT}), and compares
them to the results from the diagonalization of $\mathbf{\widetilde H}$. 
This is done for two different nuclear interactions, CD-Bonn 
\cite{machleidt01a} and AV18 \cite{wiringa95}, using a Jacobi basis.
This basis is obtained from the eigenstates of the antisymmetrizer 
(see \cite{navratil00a}). The antisymmetrized states have a well 
defined total quantum number $N$. Within each group of 
equal $N$, the states are ordered arbitrarily.  
Terms acting on the CM do not contribute in this case.
It is seen that the perturbative calculations 
can greatly improve small $\widetilde{N}_m$ results, regardless of the 
choice of the nuclear interaction. 
Figure \ref{fig:aspt} also indicates that the third-order term in 
Eq.~(\ref{eq:Ei3PT}) is small for $\widetilde{N}_m > 12$. This 
suggests that the non-diagonal matrix elements of $\widehat{\mathbf{H}}$, 
i.e., $\mathbf{C}$, are small relative to their associated energy 
denominators in second-order perturbation theory.

\subsection{$E_0$ vs. $\Delta E^2$ Scaling}
\label{sec:scalpert}
\begin{figure}
\centering
\includegraphics[width = 3.2in]{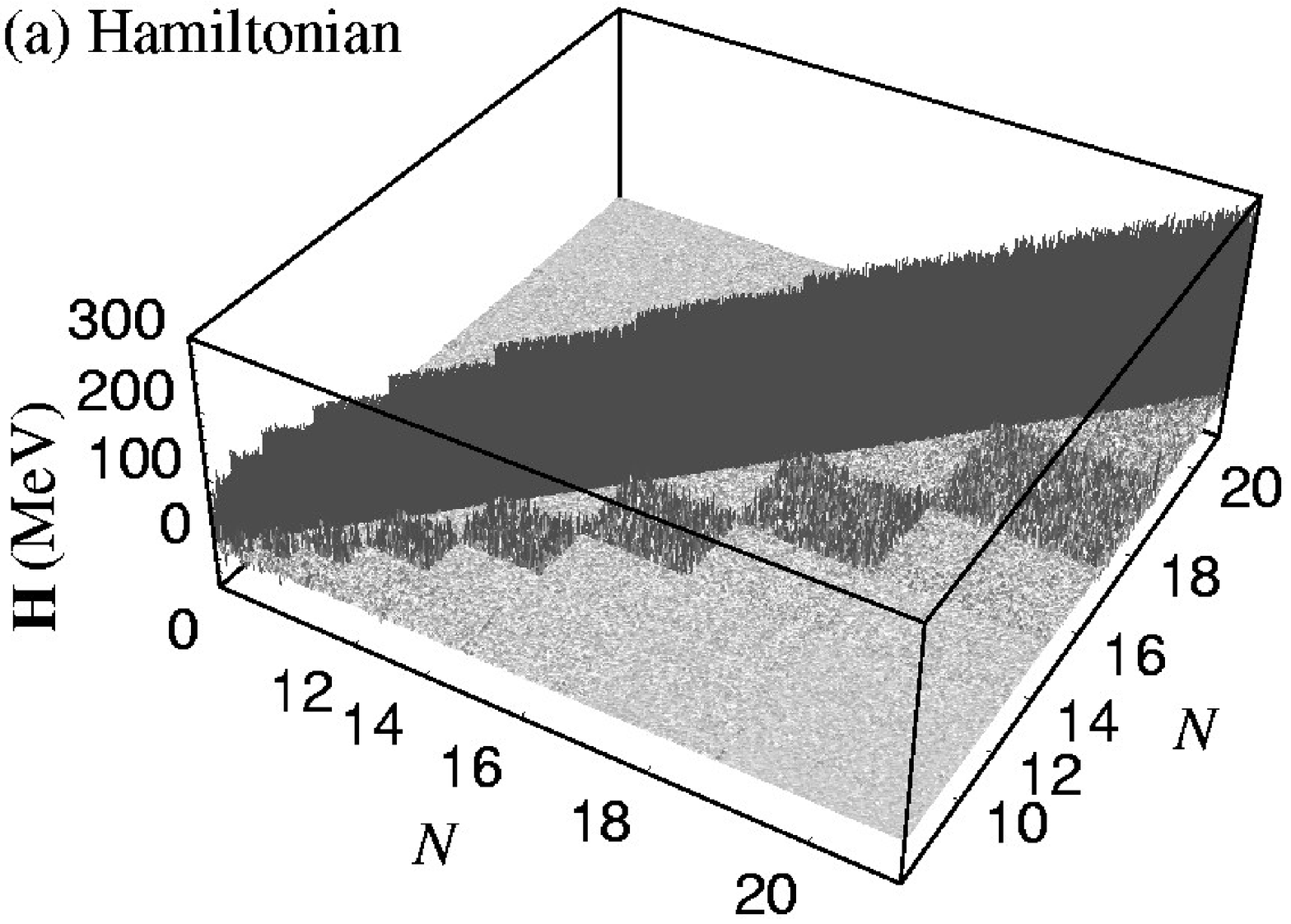}
\includegraphics[width = 3.2in]{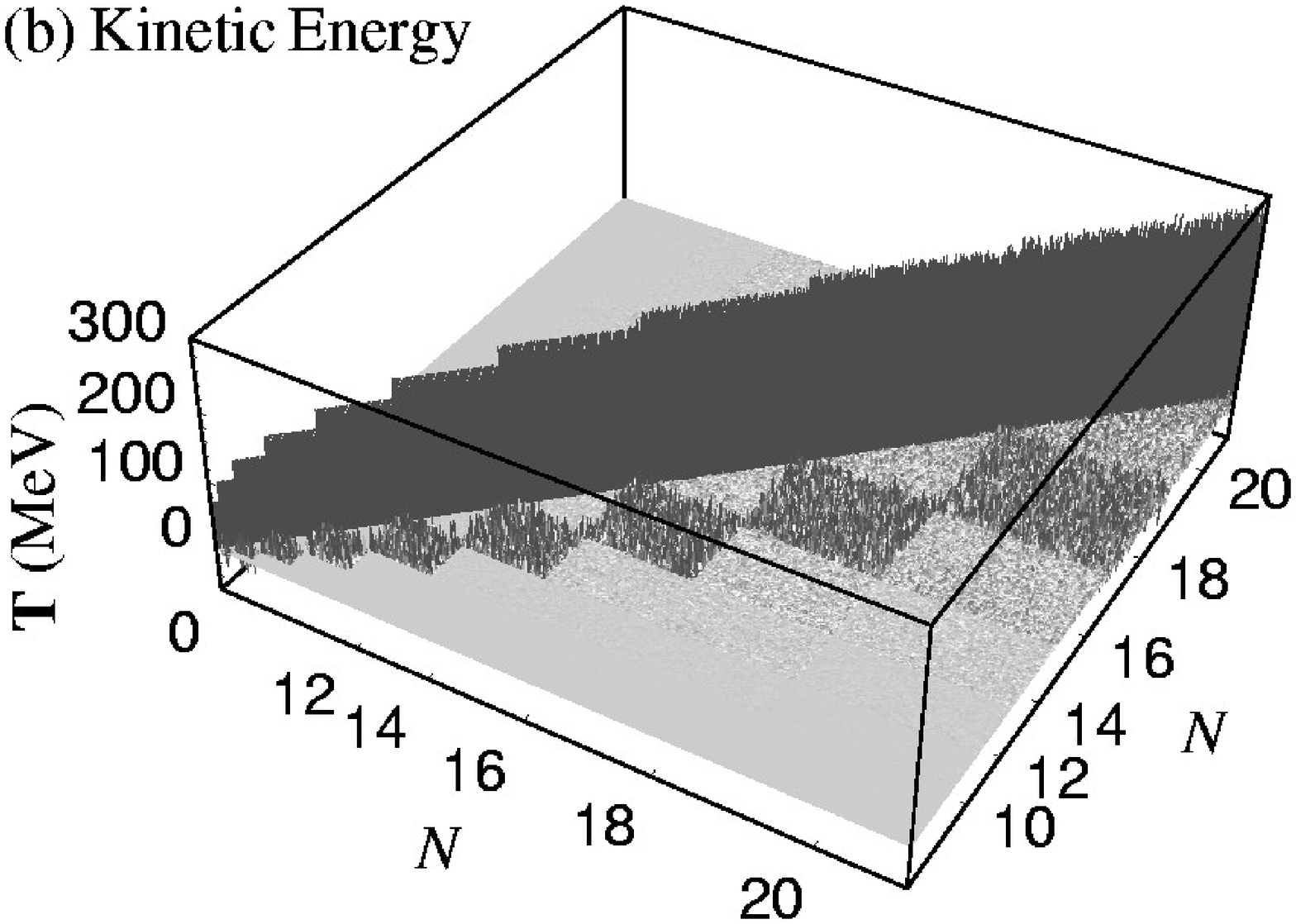}
\caption{(a) The effective Hamiltonian of the triton with 
the CD-Bonn potential and $\hbar\Omega=28$ MeV. 
The basis is sorted according 
to the quantum number $N$, but the ordering within each group of states 
with the same $N$ is arbitrary. 
This ordering will be assumed in all the following figures. 
(b) The effective kinetic energy, which is calculated in the same way 
as the Hamiltonian, but with the NN interaction turned off.
\label{fig:ash}}
\end{figure}

The expected scaling behavior can be motivated by considering the
features and results of equations (\ref{eq:Ei2PT}) and 
(\ref{eq:Ei3PT}). To this aim, the behavior of $D_{jj}$ has to be
understood in more detail. 
Figure~\ref{fig:ash} shows the matrix elements of the  
effective two-body Hamiltonian and the effective two-body kinetic energy 
for our triton model both in the 3N basis. 
The effective kinetic energy is calculated in the
same way as is the effective Hamiltonian, except that the NN 
interaction is turned off. In this case, the CD-Bonn 
potential is used with $N_m=20$ and $\hbar\Omega=28$ MeV. 
Comparing the kinetic energy with the full Hamiltonian, one sees that
the kinetic energy dominates over the contributions from the NN interaction, 
especially for large HO quantum numbers. In the figure, the basis states 
are ordered according to their total HO quantum number $N$. The ordering 
within each of these groups is arbitrary. 

\begin{figure}
\centering
\includegraphics[width = 3.2in]{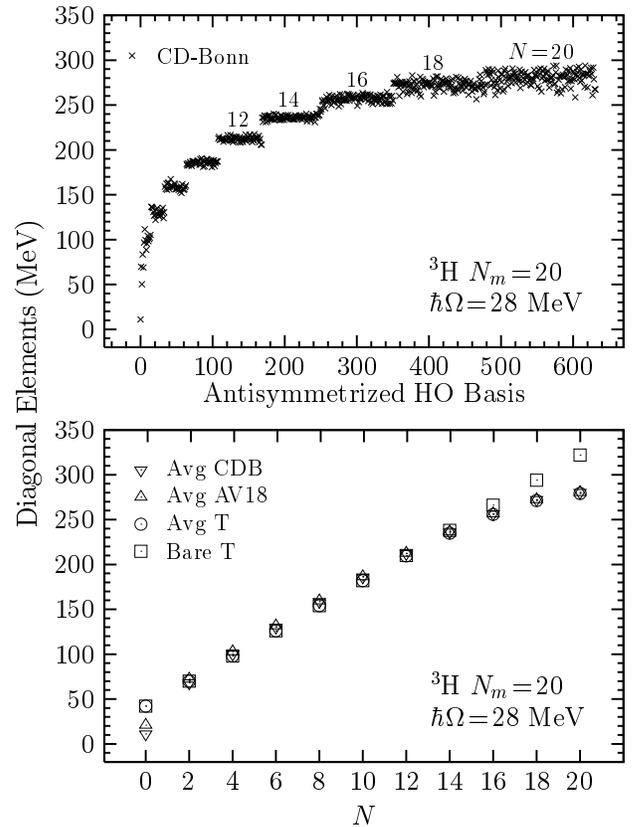}
\caption{The diagonal elements of the Hamiltonians and kinetic energies
for $^3$H. The horizontal axis in the upper panel enumerates the basis states 
for each individual diagonal element (crosses) of the model with the CD-Bonn 
potential. The horizontal axis in the lower panel indicates the HO quantum 
number $N$ for other data. Bare T (squares) is the expectation value 
of the kinetic energy operator with respect to HO basis states. The 
remaining results are the averages of diagonal elements with the same HO
quantum number $N$. From right to left, 
the first step of the individual diagonal elements in the upper panel 
corresponds to $N=20$, the next $N=18$, and so on. 
\label{fig:diags}}
\end{figure}

The potential energy affects mostly the 
low-$N$ states and the 
cross terms among them and high-lying states. The diagonal elements of the 
Hamiltonian are dominant over off-diagonal elements, which confirms our
expectation. It is also striking that a large number 
of the diagonal elements are roughly equal. 
The dark blocks next to the diagonals of the Hamiltonian and the kinetic 
energy are due to the fact that the kinetic energy operator changes the
quantum number $N$ by $\pm 2$ (also 0, which corresponds to the 
low-amplitude diagonal blocks). Since everything is renormalized to
effectively include the higher-space ($N>N_m$) influence, 
$N\pm 4, \pm6, \ldots$ terms show up in the Hamiltonian and the kinetic 
energy, and they become progressively weaker towards low $N$.
The above properties persist for other interactions, and other values of 
$\Omega$ 
or $N_m$. They are a common feature of NCSM effective Hamiltonians,
at least for the 3N system.

\begin{figure}
\centering
\includegraphics[width = 3.2in]{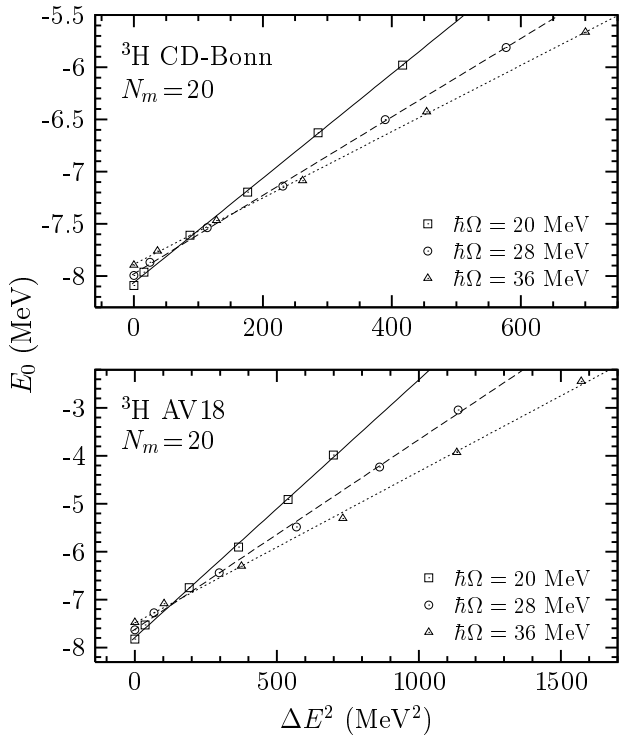}
\caption{The linear relation between $E_0$ and $\Delta E^2$. The upper
panel is for $^3$H with the CD-Bonn potential, the lower  one 
for AV18. Symbols from right
to left correspond to
$\widetilde{N}_m = 10,12,\ldots,20$ for each model. 
The lines fit 
the results from $\widetilde{N}_m = 10$ to $18$.
\label{fig:extrap}}
\end{figure}

The behavior of the diagonal matrix elements is quantified in 
Fig.~\ref{fig:diags}. The upper panel shows individual diagonal 
elements, and, again, the basis states are ordered according their HO
quantum number. The lower panel plots the averages of the diagonal 
elements that have equal $N$.    
The step structure among the individual diagonal elements reflects the fact 
that $D_{jj}$ is roughly the same in each group of states with the same 
HO quantum number. The reason for the flattening
for larger $N$ is two-fold. First, the number of states increases dramatically
for larger $N$ and, second, the renormalization of the effective matrix
elements reduces large $N$ diagonal elements. 
This is seen by comparing the average effective kinetic energy (circles)
with the bare kinetic energy (squares), which is the expectation value of the
kinetic energy operator with respect to HO basis states. For $N \le 14$, 
both follow the linear ${ (N+3)\over 2} \hbar\Omega$ behavior as expected. For
higher $N$, the effective kinetic energy turns flatter. The figure also
compares the averaged diagonal elements of the full Hamiltonians (triangles) to
the ones of the kinetic energy. This again demonstrates the dominance of the
kinetic energy for $N>0$. One can expect that the same behavior holds for more
complex nuclei.

This structure of the effective Hamiltonian guarantees that
the denominators of the second-order term in equations 
(\ref{eq:Ei2PT}) and (\ref{eq:Ei3PT}) 
are, to a high accuracy, equal over a wide range of 
$N$. Thus, we have, approximately,
\begin{equation}
\label{eq:linreg}
\mathcal{E}_0 \simeq E_0 - \frac{1}{\alpha}
\sum_{j = 0}^{\hat{n}-1} \left(\psi_0^\mathrm{T}B_j\right)^2 =
E_0 - \frac{1}{\alpha} \psi_0^\mathrm{T}\mathbf{BB}^\mathrm{T}\psi_0,
\end{equation}
where $\alpha$ is a positive constant. Since the absolute value of $E_0$
is much smaller than $D_{jj}$, the constant
$\alpha$ should be roughly the average value of the diagonal elements 
with $N > \widetilde{N}_m$. 
This also means that $\alpha^{-1}$ is only weakly dependent on
$E_0$. Neglecting higher-order perturbative terms and taking a 
constant true binding energy $\mathcal{E}_0$, it follows that 
$E_0 \propto \Delta E^2$.
This motivates a linear scaling behavior between 
$E_0$ and $\Delta E^2$.

\begin{table}
\caption{The ground state energies, $E_0$, 
of $\widetilde{\mathbf H}$ for $^3$H, 
their deviation from the correct
result $\delta E$, and the computational time necessary for the solution. 
The results of the small space ($\widetilde{\mathbf{H}}$) solution are
compared to the extrapolation method and 
the perturbative calculation using Eq.~(\ref{eq:Ei3PT}). All results are 
for CD-Bonn potential and $\hbar\Omega = 28$ MeV.
\label{tab:extrap}}
\centering
\begin{ruledtabular}
\begin{tabular}{ccccccccccccc}
 & & \multicolumn{3}{c}{$\widetilde{\mathbf{H}}$} 
 & & \multicolumn{3}{c}{Extrapolation} 
 & & \multicolumn{3}{c}{Perturbation} \\
$\widetilde{N}_m$ & & $E_0$\footnotemark[1] & $\delta E$\footnotemark[2] & 
	$t$\footnotemark[3] & & $E_{0}$\footnotemark[1] &
 	$\delta E$\footnotemark[2]\footnotemark[4] & 
	$t$\footnotemark[3] & & $E_{0}$\footnotemark[1] &
	$\delta E$\footnotemark[2] & $t$\footnotemark[3] \\ \hline
12 & & -6.502 & 1493 & 8.2 & & -7.940 &  55 &  13 & & -7.793 & 202 &  12 \\
14 & & -7.140 &  855 &  25 & & -8.015 & -20 &  39 & & -7.906 &  89 &  28 \\
16 & & -7.536 &  459 &  82 & & -7.980 &  15 & 120 & & -7.950 &  45 &  84 \\
18 & & -7.869 &  126 & 210 & & -7.970 &  25 & 340 & & -7.968 &  27 & 210 \\
20 & & -7.995 &    0 & 510 & &   --   &  -- &  -- & &   --   &  -- &  -- \\
\end{tabular}
\end{ruledtabular}
\footnotetext[1]{In units of MeV.}
\footnotetext[2]{$\delta E = E_0 - \mathcal{E}_0$, in units of keV.}
\footnotetext[3]{In units of $t_8$, the time needed to solve the ground 
state of $\widetilde{\mathbf{H}}$ with $\widetilde{N}_m = 8$.}
\footnotetext[4]{The extrapolation uses $E_0$ and $\Delta E^2$ of 
$\widetilde{\mathbf{H}}({\widetilde{N}})$ with $\widetilde{N}$ from 
10 to $\widetilde{N}_m$.}
\end{table}

\section{Application of the Extrapolation}
\label{sec:extrap}

Now that the linear scaling between $E_0$ and $\Delta E^2$ has been motivated, 
one can estimate the true binding energy $\mathcal{E}_0$
by a linear regression of $E_0$ and $\Delta E^2$ which are 
calculated with $\widetilde{N}_m<N_m$, and extrapolating it to the 
point where $\Delta E^2=0$ to estimate $\mathcal{E}_0$ at $N_m$. 

The numerical results for the relation between $E_0$ and $\Delta E^2$
are shown in Fig.~\ref{fig:extrap} for $^3$H with different NN interactions
and different values of $\hbar \Omega$.  $N_m$ is 20 in all cases, and 
$\widetilde N_m$ varies between 10 and
20. Additionally, linear fits to the results for $\widetilde N_m=10$ 
to 18 are plotted. Clearly, the linear scaling behavior is 
confirmed by these calculations. 

\begin{table}
\caption{Errors in perturbative estimates of the $^3$H binding energy 
using equations (\ref{eq:EiPT}) and (\ref{eq:Ei2PT}), and
the difference between extrapolated binding energies and corresponding
perturbative results from equation (\ref{eq:Ei2PT}). All results are for
CD-Bonn potential and $\hbar\Omega = 28$ MeV. 
\label{tab:errNm}}
\centering
\begin{ruledtabular}
\begin{tabular}{cccc}
$N_m$ & $\delta E$(Eq.~[\ref{eq:EiPT}])\footnotemark[1]  &
	$\delta E$(Eq.~[\ref{eq:Ei2PT}])\footnotemark[1]  &
	$\mathcal{E}_{\rm 0,e} - \mathcal{E}_0^{(2)}$\footnotemark[1]\footnotemark[2]  \\ \hline
 8 & 87 & 119 & -93 \\
10 & 84 & 167 & -12 \\
12 & 40 &  60 & -17 \\
14 & 52 & 103 &  3 \\
16 & 27 &  53 & -12 \\
18 & 29 &  64 &  3 \\
20 & 13 &  39 &  -7 \\
\end{tabular}
\end{ruledtabular}
\footnotetext[1]{Perturbative estimates are calculated with 
$\widetilde{N}_m = N_m - 2$. The results are in units of keV. }
\footnotetext[2]{$\mathcal{E}_{\rm 0,e}$ is the extrapolated binding energy 
using only the results of $E_0$ and $\Delta E^2$ from 
$\widetilde{N}_m = N_m - 2$ and $N_m -4$.}
\end{table}

\begin{figure}[b]
\centering
\includegraphics[width = 3.2in]{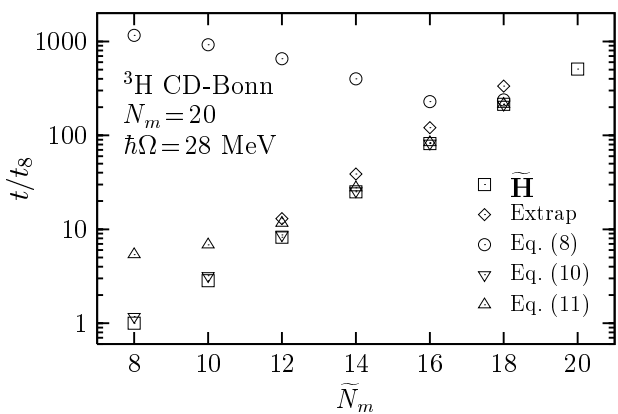}
\caption{The time consumed in the calculation in units of 
$t_8$, the time needed to solve the ground 
state of $\widetilde{\mathbf{H}}$ with $\widetilde{N}_m = 8$. 
The diamonds mark the time it takes 
to extrapolate the ground state of $\mathbf{H}$ using all 
the ground states of $\widetilde{\mathbf{H}}({\widetilde{N}})$ with
$\widetilde{N} = 10\ldots\widetilde{N}_m$, so it is the accumulation 
of the time needed to solve $\widetilde{\mathbf{H}}({\widetilde{N}})$. 
All other symbols are as in Fig.~\ref{fig:aspt}.
\label{fig:timing}}
\end{figure}

To be more quantitative, we compare results of direct calculations, the
extrapolation based on the scaling behavior, and perturbative estimates based
on Eq.~(\ref{eq:Ei3PT}) in Table~\ref{tab:extrap}.
The table indicates the computational efforts necessary by giving run
times for different calculations.
One sees that a stable extrapolation is possible starting from $\widetilde
N_m=14$. The extrapolation error for larger calculations is comparable to
the error of the perturbative estimates, indicating that both can be traced 
back
to higher order terms in the perturbative expansion. In calculations for
$N_m=30$, we have confirmed that the range of the linear behavior is 
extended for larger $N_m$. 
This is expected because $1/D_{jj}$ is driven by the kinetic
energy and, therefore, is proportional to $1/N$ (see Fig.~\ref{fig:diags}).
Consequently, if $N$ is increased by one step (i.e. 2 units), then the
change in $1/D_{jj}$ is of order $1/N^2$, which decreases with $N$.
We also note that the constant $\alpha = 268$~MeV for CD-Bonn with 
$\hbar \Omega = 28$~MeV is comparable to the diagonal elements  $D_{jj}$
shown in Fig.~\ref{fig:diags}. 

The effect of $N_m$ is demonstrated in Table \ref{tab:errNm}, where we 
list errors in perturbative estimates of the $^3$H binding energy 
using equations (\ref{eq:EiPT}) and (\ref{eq:Ei2PT}). The two equations 
are both of the second order, and the errors tend to decrease, though 
not monotonically, as the model space increases. At the same time, Table 
\ref{tab:errNm} shows that the extrapolated binding energies converge
to the results of equation (\ref{eq:Ei2PT}). This is expected, 
because our extrapolation method, i.e., equation (\ref{eq:linreg}), is 
based on an approximation of second-order perturbation theory. The 
behavior of the perturbative calculations and extrapolations suggests 
that one can reduce the extrapolation error by increasing $N_m$.

The run times are also compared in Fig.~\ref{fig:timing}. 
It is clear that perturbation theory based on Eq.~(\ref{eq:EiPT}) does not
improve the timings because of the extra time needed to diagonalize
$\widehat{\mathbf{H}}$. The extrapolation method for $\widetilde N_m=14$ 
yields sufficiently accurate results, but 
reduces the CPU-time by a factor of 13 compared to the full calculation. A
similarly accurate perturbative calculation based on 
Eq.~(\ref{eq:Ei3PT}) is still 8 times slower, which demonstrates
the usefulness of the
extrapolation to extend the calculations beyond their current limits. 

\begin{figure}[t]
\centering
\includegraphics[width = 3.2in]{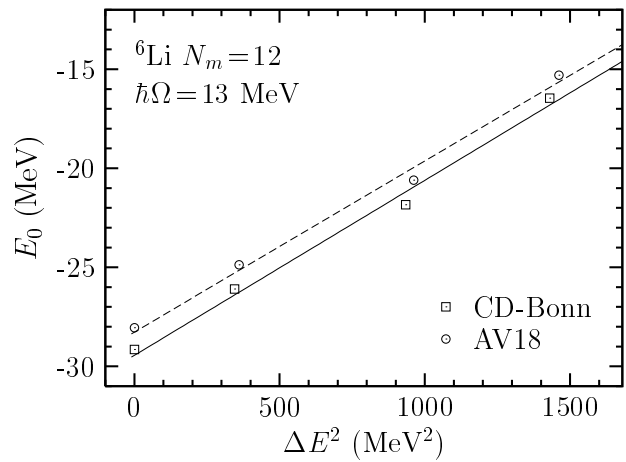}
\caption{The same as Fig.~\ref{fig:extrap}, but for the $^6$Li nucleus.
Symbols from right to left correspond to
$\widetilde{N}_m = 6, 8, 10$, and $12$ for each model. The lines fit 
the results from  $\widetilde{N}_m = 6, 8$, and $10$.
\label{fig:extrapLi}}
\end{figure}

We now apply the extrapolation method to the $^6$Li nucleus. 
The numerical results have been obtained using the Many-Fermion Dynamics 
(MFD) code \cite{varymfd}. Since we work here in a basis of slater 
determinants, we guarantee a $0s$ oscillator state of CM motion of our 
physical states by adding a CM term to equation (\ref{eq:bb}), 
$\Lambda ({\bf H}_{\rm CM} - \frac{3}{2}\hbar\Omega)$ \cite{navratil00b}, 
with $\Lambda = 10$. This separates excited 
states of CM motion from low lying physical states. 
Figure \ref{fig:extrapLi} 
shows the results with $N_m=12$ (i.e., $10\,\hbar\Omega$ above the
lowest unperturbed oscillator configuration of $^6$Li \cite{endnote24}). 
The basis dimension of the $N_m=12$ calculations is 9.7 million,
which is the largest model space published to date for $^6$Li.
This low value of $N_m$ limits what $\widetilde{N}_m$ values one
can use in the extrapolation for two reasons.
Firstly, the potential energy is not negligible at small quantum 
numbers, so $\widetilde{N}_m$ should be at least 
greater than 4 (the ground 
state of $^6$Li has $N=2$). Secondly, the accuracy of the 
extrapolation depends on the dominance of the diagonal elements of 
$\mathbf{H}$, which is also weakened at small $N$. Thus, $E_0$ and 
$\Delta E^2$ from $\widetilde{N}_m=4$ calculations are not 
in line with those from $\widetilde{N}_m=6,8,10$, and 12. 
The extrapolation errors 
are 290 keV (CD-Bonn) and 220 keV (AV18). The CD-Bonn 
potential provides 1.1 MeV more binding than the AV18 potential. 
Thus, it is interesting to 
note that the difference between the two potentials is larger than the 
extrapolation error.

\begin{figure}
\centering
\includegraphics[width = 3.2in]{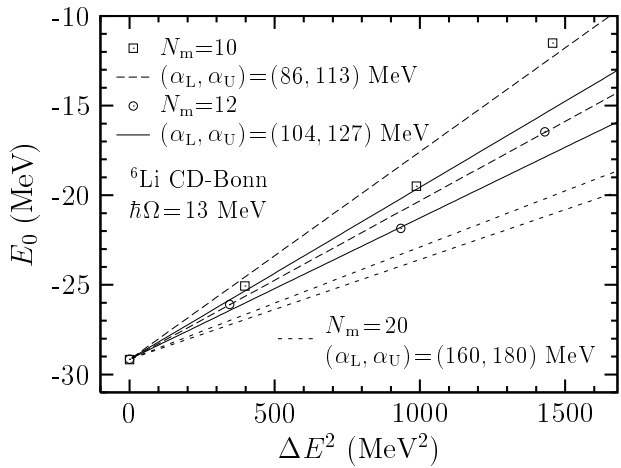}
\caption{The effect of $N_m$ on the extrapolation error. 
Symbols from right to left correspond to
$\widetilde{N}_m = N_m - 6, \ldots, N_m$. The lines follow 
$E_0 = \alpha_{\rm L}^{-1}\Delta E^2 + \mathcal{E}_0$ and
$E_0 = \alpha_{\rm U}^{-1}\Delta E^2 + \mathcal{E}_0$ for each $N_m$.
The lines with $N_m = 20$ are estimated from the effective kinetic 
energies.
\label{fig:Nm1012}}
\end{figure}

The range of $\widetilde{N}_m$ that is suitable for extrapolation 
is expected to increase with $N_m$. This is confirmed in 
Fig.~\ref{fig:Nm1012}, where an additional $N_m=10$ calculation is
made for comparison. It is seen that the results are much more linear 
for the case $N_m = 12$. Figure \ref{fig:Nm1012} 
also demonstrates why a higher value of $N_m$ is 
likely to produce a smaller extrapolation error. We can estimate
the lower and upper bounds of the constant $\alpha$, 
$(\alpha_{\rm L}, \alpha_{\rm U})$,
 in equation (\ref{eq:linreg}) using effective
kinetic energies and unperturbed ground state energies $E_0$ for 
$\widetilde{N}_m=N_m - 2$ and $N_m - 6$. The results of
$\widetilde{N}_m$ calculations are consistent with
second-order perturbation theory, because they are roughly bounded
by $\alpha_{\rm L}$ and $\alpha_{\rm U}$ for each $N_m$. 
The opening angle between the lines of slope $\alpha_{\rm L}^{-1}$ and 
$\alpha_{\rm U}^{-1}$ decreases as $N_m$ increases. 
Hence, we expect to reduce the extrapolation error by 
increasing $N_m$. 

\section{Conclusions and Outlook}
\label{sec:concl}
Because of the need for a large basis space to achieve accurate 
results, even light nuclei require a significant amount of computing 
resources to be investigated in the NCSM. 
We have justified and verified an extrapolation method for 
NCSM calculations. It is reliable, and can provide good estimates 
of large-space results from several small-space calculations. 
Sometimes, it may be
the only means for getting a useful estimate of the NCSM result for 
otherwise unachievable large model spaces.

The extrapolation formula proposed in Ref.~\cite{mizusaki03} 
agrees in leading order with our Eq.~(\ref{eq:linreg}). We would like to 
emphasize that the linear scaling between $E_0$ and $\Delta E^2$ 
is based on perturbation theory. It is not an expansion in terms of
$\Delta E^2 E^{-2}_0$, because $\Delta E^2$ can be much greater than
$E^2_0$ in the NCSM (see Fig.~\ref{fig:extrap}). 
The reasoning behind our extrapolation method 
is probably applicable only to NCSM calculations, 
because we have explicitly made use of the structure 
of the Hamiltonian in our derivation. The linear scaling between 
$E_0$ and $\Delta E^2$ relies on the flattening of the diagonal 
elements of the effective Hamiltonian (dominated by the kinetic 
energy) as $N$ approaches $N_m$. With this behavior, the energy 
denominator in equation (\ref{eq:Ei2PT}) can be approximated by a 
constant $\alpha$. 

Generally speaking, extrapolation 
methods depend not only on the structure of the Hamiltonian but also 
on the truncation scheme that is used to produce an approximate state. 
Different truncation schemes may lead to different scaling behavior 
\cite{mizusaki03}. The traditional phenomenological shell model does 
not generate the 
structure of the Hamiltonian that is advantageous for our method. 
Specifically, in calculations with a core diagonal dominance is 
reduced since energies relative to a core are obtained. For realistic 
mean field potentials, the single particle spectrum does not rise as 
fast as an oscillator spectrum which itself does not rise as fast as 
the kinetic spectrum in an oscillator basis. Hence, the diagonal 
dominance we have in the NCSM is much stronger than the traditional 
shell model, and our method probably cannot be applied to the 
traditional shell model without modifications. On the other hand, the 
methods for the traditional shell model do not necessarily
apply to the NCSM either. In fact, it is evident from 
Fig.~\ref{fig:extrapLi} that a quadratic fit \cite{mizusaki03} to the 
$\widetilde{N}_m=6$, 8, and 10 results will yield a significantly 
larger error.

From Table \ref{tab:errNm} we have learned that for small model spaces 
there are two competing sources of inaccuracy: one in the perturbation 
theory result and the other in the extrapolated result. We have shown 
that the extrapolated result is converging to the perturbation theory 
result, equation (\ref{eq:Ei2PT}). In Table \ref{tab:errNm} one sees 
that the deviation of equation (\ref{eq:Ei2PT}) and the extrapolation 
is already small even for relatively small values of $N_m$, e.g., 
$N_m = 10$. Since perturbation theory requires larger $N_m$, around 16, 
to give estimates of similar accuracy, we can conclude that overall our 
results are probably dominated by errors due to perturbation theory. 

This method, like NCSM calculations themselves, is limited by the size 
of the model-space, because $\Delta E^2$ has to be evaluated in the 
full $P$ space. Nevertheless, it has been shown to be a valuable tool.
Its power is based on the much smaller dimension of the $\mathbf{B}$ 
matrix compared to the full matrix for the Hamiltonian operator. 
Calculations for even larger model spaces will become possible once we 
make explicit use of the small dimension of the $\mathbf{B}$ matrix in 
our codes.

In Figs.~\ref{fig:extrap} and \ref{fig:extrapLi},
we observe that the extrapolation error, i.e., 
$\delta E = E_0 - \mathcal{E}_0$, at $\Delta E^2=0$ is quite small
compared with differences in exact results of $\mathcal{E}_0$ for 
various values of
$\hbar\Omega$ or choices of potentials. Therefore, the extrapolation can 
also serve as a way to estimate the uncertainties of NCSM
results arising from $\hbar \Omega$ dependence and choices of 
interactions. This will be an important application of our method  in 
future investigations of nuclei within the NCSM. 

\section*{Acknowledgments}
H.Z., A.N. and B.R.B. acknowledge partial support by NSF grant 
No.~PHY0070858. J.P.V. acknowledges partial support by USDOE grant 
No.~DE-FG-02 87ER40371. This work was partly performed under the 
auspices of the U.S. Department of Energy by the University of 
California, Lawrence Livermore National Laboratory under contract
No. W-7405-Eng-48. This research used 
resources of the National Energy Research Scientific Computing Center, 
which is supported by the Office of Science of the U.S. Department of 
Energy under Contract No. DE-AC03-76SF00098.

\end{document}